# Towards ultrasensitive biosensors based on virus-like particles and plasmonic surface lattice resonance

Weronika Andrzejewska,[a,*] Nadzeya Khinevich,[b] Patryk Obstarczyk,[c] Szymon Murawka,[a] Tomas Tamulevičius,[b] Joanna Olesiak-Bańska,[c] Sigitas Tamulevičius[b] and Mikołaj Lewandowski[a,*]

[a]*NanoBioMedical Centre, Adam Mickiewicz University, Wszechnicy Piastowskiej 3, 61-614 Poznań, Poland*

[b]*Institute of Materials Science, Kaunas University of Technology, K. Baršausko St. 59, 51423, Kaunas, Lithuania*

[c]*Institute of Advanced Materials, Wroclaw University of Science and Technology, Wybrzeże Wyspiańskiego 27, 50-370 Wrocław, Poland*

*\*Corresponding authors: weronika.andrzejewska@amu.edu.pl, lewandowski@amu.edu.pl*

*ORCiD: 0000-0002-8565-3099 (W.A.), 0000-0001-9348-3918 (N.K.), 0000-0001-9911-3301 (P.O.), 0000-0002-6134-0813 (S.M.),0000-0003-3879-2253 (T.T.), 0000-0002-7226-0077 (J.O.-B.), 0000-0002-9965-2724 (S.T.), 0000-0002-1480-8516 (M.L.)*

## Abstract

Plasmonic surface lattice resonance (SLR) is a phenomenon in which individual localized surface plasmon resonances (LSPRs) excited in periodically-arranged plasmonic nanoparticles couple through the interaction with propagating diffracted incident light. The SLR optical absorption peak is by at least one order of magnitude more intense than the LSPR one, making SLR superior for applications in which LSPR is commonly used. Recently, we have developed a route for the fabrication of spherical virus-like particles (VLPs) with plasmonic cores and protein coronas, where the LSPR in the cores amplifies vibrational signals originating from protein-antibody bonding, showing the potential of VLPs in biodetection. However, the signals were not strong enough to detect antibodies at very low concentrations. Here, we show that by ordering the VLPs in periodic nanoarrays exhibiting SLR amplifies the signals by two orders of magnitude, revealing superior potential of SLR arrays in ultrasensitive biodetection.

1. Introduction

Biosensing is one of the most developing areas of science and technology. Many diseases, including viral and bacterial infections, as well as cancer, are manifested by the appearance of specific biomolecules in the human body. Detection of these molecules at the very early stage of the disease, i.e. when the molecules are in extremely low concentrations, is crucial for proper diagnosis and effective treatment. Currently, tremendous efforts are being made to invent reliable and easy-to-use medical tests that would provide both unprecedented sensitivity and rapid detection time for different viruses, bacteria, proteins, antibodies, antigens, nucleic acids and other biomolecules [1–3]. Modern biosensors are based on sophisticated biological, physical and chemical phenomena, and require the use of state-of-the-art technological equipment for their fabrication. One of the most innovative solutions in the field of optical sensors includes utilizing the so-called localized surface plasmon resonance (LSPR) [2]. LSPR is a phenomenon related to the collective oscillation of free electrons at the interface of noble metal nanostructures and the surrounding dielectric medium upon absorption of light of a specific wavelength [4]. As the resonance wavelength depends on the dielectric environment of plasmonic nanoobjects, shifts of the resonance absorption peak are often used as indicators of biomolecules adsorption. However, this kind of biodetection lacks specificity necessary for unambiguous diagnosis. A more specific approach is related to the detection of vibrational signals characteristic of specific biomolecules. In this respect, the electric field generated by the LSPR can be used to amplify Raman signals coming from species located in the vicinity of plasmonic nanostructures – a phenomenon known as surface-enhanced Raman spectroscopy (SERS).

It was relatively recently discovered, that by arranging plasmonic species in ordered periodic arrays – with the distance between the objects being adjusted to the wavelength of the incident light – a far-field coupling of the scattered radiation fields of neighboring nanostructures is generated in the lattice plane, giving rise to the so-called plasmonic surface lattice resonance (SLR) [5,6]. The position of the optical absorption peak of SLR depends on the size of plasmonic objects and the periodicity of the matrix. Notably, the intensity of the SLR signal is by at least an order of magnitude more intense and exhibits higher quality factor than the SPR one [7]. Moreover, its position may be easily tailored to a required wavelength [8,9]. All these make SLR arrays superior for applications in which SPR is commonly used, such as optoelectronics, photovoltaics, biosensing and SERS [10–12].

In our recent work, we have presented the route for fabricating spherical virus-like particles (VLPs) imitating the SARS-CoV-2 virus and consisting of plasmonic Au cores and S1-protein coronas [13]. The LSPR in the cores was found to amplify vibrational signals originating from the bonding between the S proteins and the anti-S monoclonal antibodies (mAbs), showing the potential of fabricated VLPs in biodetection. However, the signals were too weak to detect the antibodies at very low concentrations. In this Letter, we show that by ordering the VLPs into a periodic array using the capillary-assisted particle assembly (CAPA) technique [14,15] it is possible to further amplify the signals approx. 350 times and, thus, obtain ultrasensitive detection of viral antibodies. The amplification is related to the presence of SLR in the studied system and the associated SERS. It must be emphasized, that in a typical CAPA process the deposited objects are being suspended in an inorganic solvent, e.g. ethanol. However, these kinds of solvents are not suitable for biomolecules which are supposed to remain biologically active. Thus, the whole process had to be optimized using the water solvent and dispersants. To the best of our knowledge, this is the first work reporting the fabrication of such a matrix consisting not of bare noble metal particles, but particles consisting of noble metal cores and protein coronas, which opens a route for the development of a whole new family of biosensors based on VLPs and SLR.

2. Materials and Methods

The SARS-CoV-2 VLPs were prepared according to the protocol described in Ref. [16]. In brief, citrate-stabilized 10 nm nanoparticles (NPs) were obtained using the Turkevich method and subsequently overgrown with gold in the presence of cetyltrimethylammonium bromide (CTAB) to $90 < d \leq 100$ nm. Next, the solution was centrifuged, the supernatant was discarded several times, the CTAB was replaced with bis(p-sulfonatophenyl)phenylphosphine dihydrate dipotassium salt (BSPP) and the particles were resuspended in Milli-Q water. Finally, purified recombinant SARS-CoV-2 S1 protein in concentration of 6.25 µg/mL was added to BSPP-coated AuNPs (109 particles/mL) and the mixture was incubated for 1 h at room temperature.

The VLPs were deposited onto structured polydimethylsiloxane (PDMS) substrates using a CAPA setup equipped with a PI miCos LS-110 motorized linear precision translator, Meerstetter Engineering TEC-1090 temperature control system, Olympus BX51 optical microscope and a QImaging Micropublisher 3.3 CCD camera. The substrates featured square-arranged round holes spaced every 420 nm obtained by using a silicon mold fabricated with soft lithography. VLPs were dispersed in water with physiological pH to maintain the natural conformation and biological activity of the protein corona. Then, 100 µl was drop-cast onto PDMS placed on the moving stage and covered with the fixed microscope slide forming a meniscus. Finally, the drop was moved across the surface with a velocity of 1 µm/s, at a 20.8°C dew point and under ambient pressure conditions.

The fabricated systems were studied with respect to their structure, optical and vibrational properties. The transmission electron microscopy (TEM) measurements were performed using a JEOL JEM-1400 microscope operating at 120 kV. 10 µl of VLPs were dried for 1 min on a standard 300-mesh Cu grid with carbon Formvar (Agar Scientific) and the remaining liquid was removed by touching the grid edge with a low lint paper. The samples were stained with 5 µl of 2% uranyl formate solution, which was removed after 1 min. Scanning electron microscopy (SEM) studies were carried out with FEI Quanta 200 FEG instrument under low vacuum and analyzed using the ImageJ software [25]. The electron gun voltage was 10 kV, WD 8.6-8.7 mm and the spot size was set to "3". Atomic force microscopy (AFM) images were obtained using the Bruker Dimension Icon microscope working in the ScanAsyst measurement mode and analysed using the Gwyddion 2.56 software [26]. UV-Vis-NIR spectra were recorded using an Olympus optical microscope equipped with a fiber-optic spectrometer Avantes AvaSpec-2048 with 1.4 nm resolution. Raman spectra were collected with a Renishaw Raman spectrometer using 532 nm wavelength excitation with 45 mW power, equipped with a Leica 50x/0.75 NA, 2400 lines/mm grating and thermoelectrically-cooled 1024 pixels CCD.

## 3. Results and Discussion

The general scheme of the experiment is presented in Figure 1. First, SARS-CoV-2 VLPs were fabricated following the protocol described in Ref. [16] (step I). The particles were then deposited using CAPA onto a PDMS substrate with periodically arranged holes (step II). For this purpose, a condensed stock water solution of VLPs was drop-casted onto a PDMS substrate with holes arranged in a square lattice with 420 nm period fabricated using a silicon mold, and the particles were directed towards the holes with a precisely positioned glass slide. The method exploits long-range capillary interactions to direct the particles to the assembly, with the capillary force ($F_c$) acting perpendicular to the meniscus formed in the particle accumulation zone and pushing the VLPs into the traps. In that way, a well-ordered array of particles was formed (step III). Finally, a solution of mAbs was drop-cast onto the substrate and Raman signals related to the protein-antibody interaction were recorded (steps IV and V).

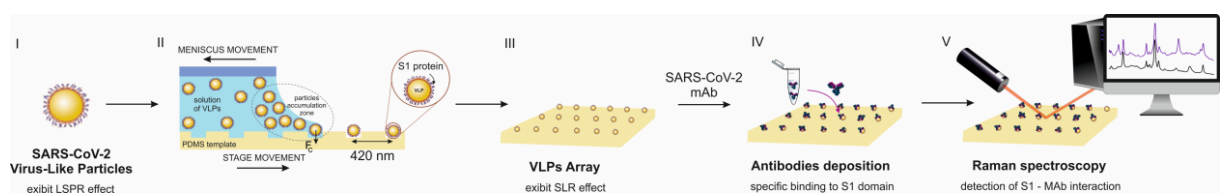

**Figure 1.** The general scheme of the experiment.

The structural characterization of fabricated systems was performed in steps I-IV using TEM, SEM and AFM. Figure 2(a) shows a TEM image of an exemplary SARS-CoV-2 VLP. It consists of a 100 nm Au NP surrounded by S1 domains of the real SARS-CoV-2 coronavirus. A detailed description of the particles is presented in Ref. [16]. SEM image of the PDMS matrix is shown in Figure 2(b). A periodic array of holes arranged in a square manner is clearly visible. Figure 2(c) presents SEM images of a CAPA-fabricated array of VLPs. The micrographs were taken at two magnifications – 50,000x and 180,000x (top and bottom row, respectively). Also, for better visualization, one more image was taken for the VLPs array at a slight angle with respect to the substrate normal (Figure 2(d)). All the micrographs revealed that VLPs indeed reside in the holes of the polymer matrix and that the coverage is uniform. Except for a few places where two particles were residing in one hole and for a few empty spots, a single VLP was observed in each hole of the PDMS substrate. To confirm the macroscopic uniformity of the fabricated VLPs nanoarray, it was transferred to a microscope glass slide and photographed on black and white backgrounds (Figure 2(e)). The red/violet color of the array visible on the image results from the absorption of light by the array and is directly related to the size of Au NPs used for VLPs preparation, as well as the periodicity of the matrix. The fact that the color was uniform across the entire PDMS substrate, i.e. 3 cm$^2$ area, confirmed the nearly 100% coverage with VLPs. Similar SEM micrographs recorded after the deposition of mAbs are shown in Figure 2(f). No significant changes in the morphology can be observed, except for the increase in the VLPs' diameter. Even after washing the array several times with distilled water, the VLPs remained in place. This indicated the stability of fabricated array.

Figure 2(g) shows an AFM image of an array of pure Au NPs, Figure 2(h) an image of a similar array of SARS-CoV-2 VLPs and Figure 2(i) an array of VLPs exposed to mAbs. The surface profiles of these samples are summarized in Figure 2(j). Pure Au NPs have a diameter of around 100 nm, VLPs of approx. 130 nm (due to attachment of S1-proteins to Au) and the particles exposed to mAbs were more than 150 nm in diameter. The results confirm that the CAPA process does not affect the S1-protein shell around the VLPs and that the mAbs may attach to the VLPs in the array.

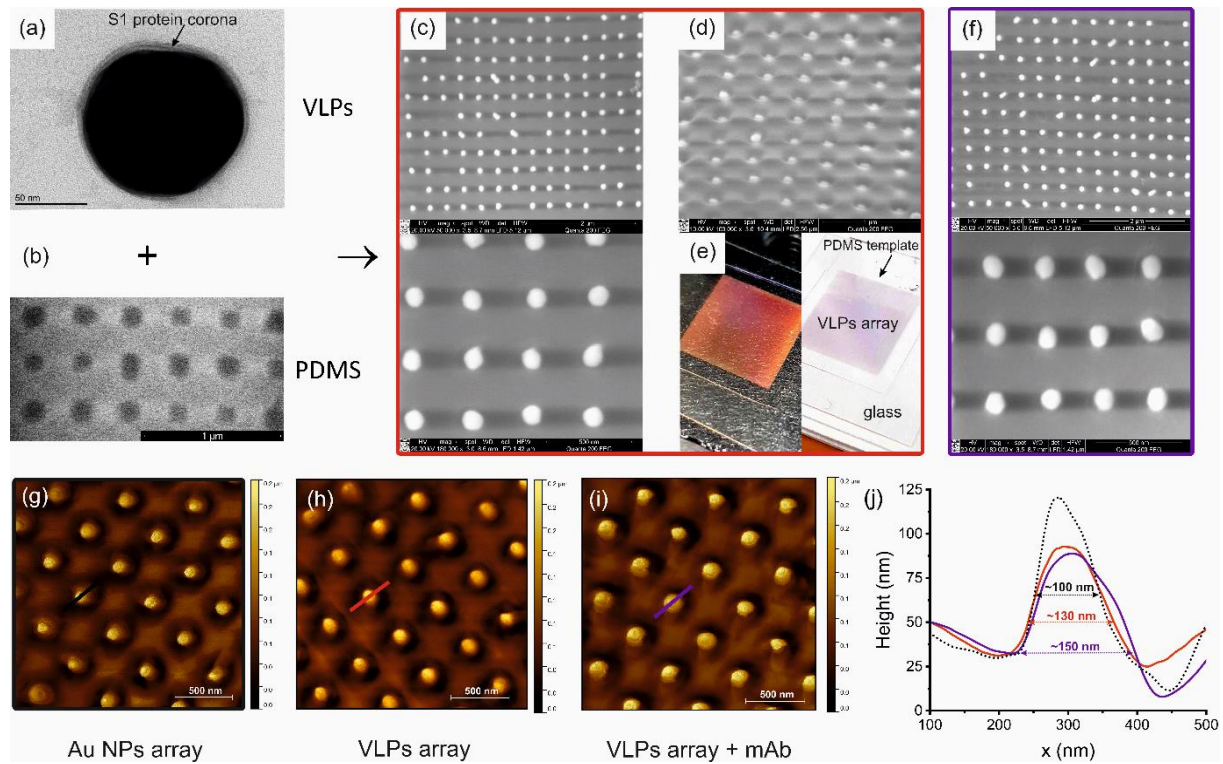

**Figure 2**. (a) A TEM image of an exemplary SARS-CoV-2 VLP with a visible S1-protein corona around the 100 nm Au core (a). (b) shows an SEM image of the PDMS substrate with holes engraved using the silicon mold, while (c) and (d) microscopic images of the VLP array fabricated using the CAPA method. The macroscopic photographs of the VLP array, taken on two different backgrounds, reveal that the coverage is uniform. (f) shows SEM images of the array following the deposition of mAbs. (g-i) show AFM images of Au NPs array (g), VLPs array (h) and VLPs+mAbs array (i). The line profiles marked on the images are presented in (j).

As already stated, periodic arrays of noble metal NPs exhibit the SLR effect, with the position of the resonance peak being dependent on the size of NPs, the periodicity of the matrix and the refractive index of the surrounding medium [17–19]. The optical properties of plasmonic matrices are usually determined with UV-Vis-NIR spectroscopy, which allows determining the position and intensity of the LSPR/SLR absorption signal [20]. Figure 3(a) shows UV-Vis spectra obtained for an array of pure Au NPs (black dashed line) and an array of VLPs before (red) and after (blue) exposure to mAbs. The position of the resonance peak observed for the array of pure Au NPs is 623 nm, which can be considered as a reference value. In the case of VLPs, the peak is located at 632 nm, which is related to the change in the electron density and refractive index of the medium surrounding the plasmonic cores (which is the S1-protein corona) [13]. After exposure to mAbs, the signal shifts by approx. 2 nm to lower wavelengths (see the inset to Figure 3(a)). This shift further confirms the successful attachment of mAbs to S1-proteins and indicates that SLR is more sensitive to the changes in the dielectric environment further from the plasmonic cores than the LSPR, for which no shift was observed following mAbs attachment [13]. Similar shifts in the resonance peak position are often used as indicators of specific binding in biodetection. However, it has to be kept in mind that a shift may also occur for a non-specific interaction or charge redistribution at the surface [21]. What is important from the point of view of the present work, the SLR signal recorded for all matrices exhibits a characteristic sharp shape, indicating the effectiveness of the CAPA process in producing well-ordered arrays. This is particularly meaningful for VLPs, which are not easy to deposit because of the use of water as a solvent and the presence of high-viscosity protein macromolecules. In a typical CAPA process, the deposited particles are suspended in inorganic solvents, e.g. ethanol. Such solvents may damage biological molecules affecting their functionality by promoting denaturation. In our case, the entire process was planned and optimized using water as a solvent, and with dispersants (CTAB and BSPP).

In order to utilize the fabricated plasmonic arrays for the specific biodetection of anti-S mAbs, we have performed SERS measurements focusing on the analysis of vibrational signals characteristic of the S1-proteins/mAbs interaction. These signals were identified during our studies in VLPs in solution [13]. The Raman spectra obtained for the VLP array before and after exposure to mAbs are shown in Figure 3(b). The line recorded for the VLPs nanoarray (red curve) shows the fingerprint bands, including those originating from the BSPP and the S1-protein, as well as bands assigned to PDMS. In particular, the presence of the S1-protein is manifested by the signal centered at around 1440 cm$^{-1}$. In this region, fingerprints of polar amino acids are observed, specifically the scissoring modes coming from CH$_2$ vibrations (1426 and 1438 cm$^{-1}$ for asparagine and 1427 and 1450 cm$^{-1}$ for glutamine) [22]. After exposure to anti-S mAbs, two additional expected bands, characteristic of protein/antibody interaction, appeared at 1332 and 1558 cm$^{-1}$ (blue curve) [13]. Thus, the result confirmed the biological activity of VLPs in the array and the attachment of mAbs to VLPs. It also has to be noted, that the intensity of recorded peaks was much higher than in the case of VLPs/mAbs in solution (black curve), even though the number of VLPs in the array was much lower than in the solution. This indicates the role of the SLR in the amplification of Raman signals. To directly determine the strengthening effect of the SLR, we have determined the concentration of particles at the surface in the case of drop-casted solution of VLPs (exhibiting the LSPR) and in the VLPs nanoarray (where LSPR and SLR occur), and then correlated that with the Raman signal intensities in both systems. A typical drop of the solution has a volume of 0.2 ml and VLPs concentration of 1.66 fM. This gives 2×10$^8$ particles in a drop that are distributed over an area of 1 mm$^2$ when drop-cast. On the VLPs array, the particles are arranged in a square lattice with 420 nm period. Assuming 1 particle per spot (on average), this gives 5.7×10$^6$ particles per 1 mm$^2$. Thus, it may be concluded that in the substrate area covered by the drop there are 35 times more particles than within the same area in an array. However, despite the much lower number of plasmonic particles the Raman signals detected from an array had roughly 10 times higher intensity than those recorded from a drop. This indicates that the SLR amplifies the signals approx. 350 times (the value could be probably even higher if the laser wavelength is adjusted to the specific material system), which allowed us detecting mAbs at 25 µM concentration. This proves the very high sensitivity of the fabricated biosensor.

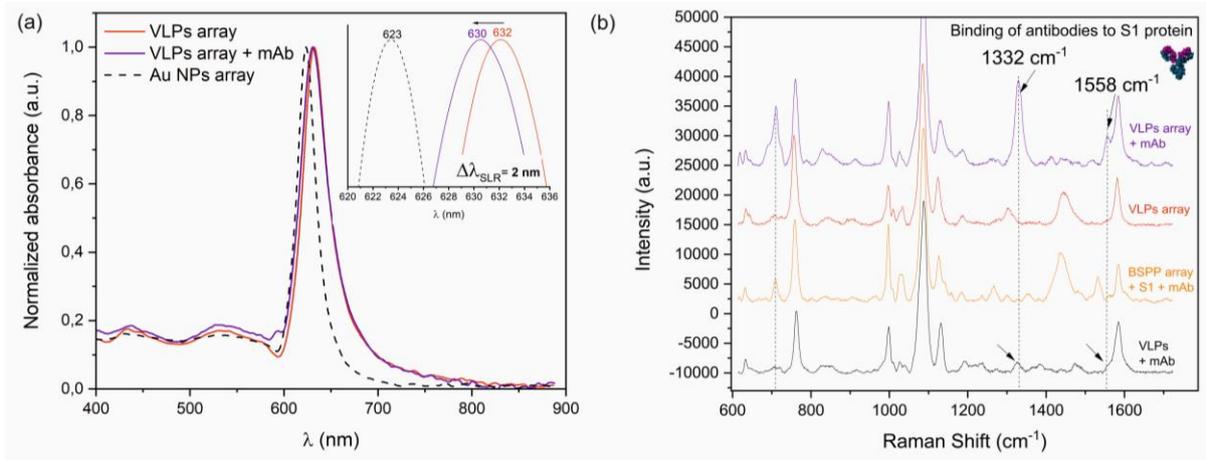

**Figure 3**. (a) UV-Vis-NIR spectra with marked SLR peak positions (inset). (b) shows Raman spectra obtained for an array of VLPs (red line), an array of VLPs with mAbs (violet line) and for VLPs with mAbs in solution (black).

Finally, we have determined whether it is possible to biofunctionalize an array of pure Au NPs on a PDMS substrate. The absence of organic molecules immobilized on NPs during the deposition process allows the use of inorganic solvents during the CAPA deposition, making the process adjustable to the materials used (template and particles). In the experiment, Au NPs were covered only with BSPP to prevent their aggregation and allow future attachment of S1-proteins. Following deposition, the solution of proteins was drop-casted onto the array to biofunctionalize the NPs. Then, a solution of mAbs was applied to confirm the usefulness of the system in biodetection. The Raman spectrum obtained after exposure to mAbs is compared with the other spectra in Figure 3 (in orange). Notably, a signal originating from the S1-proteins is visible at around 1440 cm$^{-1}$ and no signals indicating the specific interaction between mAbs and S1-proteins are observed at 1332 and 1558 cm$^{-1}$. This indicates that specific binding between the proteins and antibodies did not occur. This kind of behavior can be explained by considering the properties of the PDMS substrate. In the case of the array made of VLPs, the S1 protein is present at the surface of Au NPs in the form of dense protein corona. Also, the proteins are most probably oriented in a specific way determined by the bonding with gold. When it comes to AuNPs coated with BSPP, they were biofunctionalized when deposited onto the holes in PDMS. In such a case, protein molecules – to reach the surface of gold – must go into the holes, which is not easy due to geometric constraints and the hydrophobic nature of the polymer (the literature reports indicate that

PDMS has to be additionally functionalized in order to be suitable for biological applications [23]). Thus, the binding of proteins to Au NPs on PDMS is not as trivial and well-defined as in the case of particles in solution. In the case of mAbs, the binding could be even more difficult due to the more limited space and the hydrophobicity of PDMS mentioned above. It is worth mentioning that the Raman spectrum recorded following the deposition of mAbs shows additional peaks centered around 1268, 1353 and 1534 $cm^{-1}$, which could originate from proteins/antibodies [24]. These peaks were not observed for the VLPs-mAbs systems, which further indicates that the interaction between mAbs and S1-proteins is different than in the case of VLPs.

## 4. Conclusions

In summary, we have used the CAPA method to fabricate a well-ordered array of SARS-CoV-2 VLPs. The aim was to enhance the characteristic Raman signals originating from the specific interaction between the S1-proteins present at the surface of VLPs and the anti-S antibodies by utilizing the plasmonic SLR. The main achievements of this work are:

- the fabrication of structurally stable nanoarrays of VLPs by preserving the biological activity of proteins;
- the enhancement, by two orders of magnitude, of the intensity of vibrational signals originating from the specific protein-antibody interaction.

The results open new pathways for ultrasensitive biodetection based on combined effect of VLPs and SLR.

*Author contributions*

W.A.: conceptualization, methodology, validation, formal analysis, investigation, data curation, writing—original draft and visualization. N.K.: methodology, investigation and data curation. S.M.: AFM-investigation and data curation. P.O.: Au NPs synthesis—investigation. J.O.-B.: AuNP synthesis—resources and supervision. T.T.: CAPA—methodology and resources, SEM—investigation. S.T.: resources, supervision, project administration and funding acquisition. M.L.: conceptualization, methodology, resources, writing—original draft, writing—review and editing, supervision, project administration and funding acquisition.

*Acknowledgments*

The studies were performed within the LaSensA project carried out under the M-ERA.NET 2 scheme (European Union's Horizon 2020 research and innovation programme, grant No. 685451) and co funded by the Research Council of Lithuania (LMTLT), agreement No. S-M-ERA.NET-21-2, the National Science Centre of Poland, project No. 2020/02/Y/ST5/00086, and the Saxon State Ministry for Science, Culture and Tourism (Germany), grant No. 100577922, as well as from the tax funds on the basis of the budget passed by the Saxon State Parliament.